\documentstyle [prl,aps,multicol,amsfonts,amstex]{revtex}
\renewcommand{\narrowtext}{\begin{multicols}{2} \global\columnwidth20.5pc}
\renewcommand{\widetext}{\end{multicols} \global\columnwidth42.5pc}

\title{Exact Multifractality for Disordered $N$-Flavour Dirac  
Fermions in Two Dimensions}   

\author {Jean-S\'ebastien Caux}

\address{Department of Physics, University of Oxford, 1 Keble Road,
Oxford, OX1 3NP, UK} 

\date{$Date:$ \today}

\begin{document}

\bibliographystyle{prsty}

\draft
\maketitle

\begin{abstract}
We present a nonperturbative calculation of all multifractal
scaling exponents at strong disorder for critical wavefunctions of
Dirac fermions interacting with a non-Abelian random vector potential
in two dimensions.  The results, valid for an arbitrary number of
fermionic flavours, are obtained by deriving from Conformal Field Theory
an effective Gaussian model for the wavefunction amplitudes and
mapping to the thermodynamics of a single particle in a random
potential.  Our spectrum confirms that the wavefunctions remain 
delocalized in the presence of strong disorder.
\end{abstract}



\narrowtext For a noninteracting disordered system near a
metal-insulator transition, there now exists compelling evidence that
wavefunctions exhibit {\it multifractal} behaviour at scales below the
localization length $L_c$.  Namely, perturbative renormalization-group
analysis of replicated \cite{WegnerZPB36} and sypersymmetric
\cite{Fal'koPRB52} nonlinear sigma models as well as numerical work
(see for example \cite{JanssenIJMPB8}), point to the existence of
irregular scaling properties of the local moments of the wavefunction
distribution,
\begin{eqnarray}
Z(q) = \int \frac{d^2 x}{a^2} |\Psi(x)|^{2q} ,
\label{Z}
\end{eqnarray}
with respect to system size $L < L_c$ ($a$ is  the short-distance
cutoff).  The information regarding these irregularities is most
conveniently condensed into a scaling spectrum $\tau(q)$ defined in
terms of the above moments by
\begin{eqnarray}
\tau (q) = \lim_{a/L \rightarrow 0} \frac{1}{\ln (a/L)} \langle [\ln
Z(q) -q\ln Z(1)]\rangle,
\label{tau}
\end{eqnarray}
where the brackets denote disorder averaging over a suitable ensemble,
and $Z(1)$ plays the role of the wavefunction normalization. In this
framework, a localized state has $\tau(q>0) =0$, while $\tau(q)
\propto q$ for a simple fractal; any nonlinearity signals
multifractality.

There exists a well-studied class of exactly solvable toy models which
are known to exhibit the phenomenon of multifractality, namely Dirac
fermions subject to a random (Abelian or non-Abelian) vector potential
(RVP) in 2-d Euclidean space
\cite{HoPRB54,LudwigPRB50,KoganPRL77,ChamonPRL77,CastilloPRB56,%
MudryNPB466,NersesyanPRL72,CauxNPB466,CauxPRL80,Bernardh9509137}.  The
Abelian case
\cite{HoPRB54,LudwigPRB50,KoganPRL77,ChamonPRL77,CastilloPRB56,%
MudryNPB466} corresponds to a single flavour of 2-d Dirac fermions
interacting with a 
random magnetic field, and applies to disordered systems whose Fermi
surface is effectively made of a single node characterized by a
$V$-shaped singularity at the Fermi energy (as for example the
Chalker-Coddington network model, or degenerate semiconductors
\cite{HoPRB54}).  Wavefunctions are known
\cite{LudwigPRB50,MudryNPB466} to be localized in the vicinity of the
Fermi energy except exactly at the Fermi level, where critical states
exist.  Their full multifractal spectrum was recently constructed in
\cite{ChamonPRL77,CastilloPRB56} using a mapping to a generalized
random energy model.

On the other hand, the non-Abelian random vector potential (NARVP)
model
\cite{MudryNPB466,NersesyanPRL72,CauxNPB466,CauxPRL80,Bernardh9509137}
describes systems with a Fermi surface collapsing to an arbitrary
(even for a tight-binding Hamiltonian) number of Fermi points in the
2-d Brillouin zone.  To each node then corresponds a different flavour
of Dirac fermions.  Besides being an effective theory for low-energy
excitations of a $d$-wave superconductor \cite{NersesyanPRL72}, the
NARVP problem was recently obtained in the context of non-Hermitian
Quantum Mechanics as the effective model for a 2-d particle in a
random impurity potential subject to an imaginary (driving) vector
potential \cite{Mudry9712103}.  This last model has found applications
in a wide variety of physical systems ranging from fluctuating vortex
lines in superconductors with columnar defects \cite{HatanoPRL77} to
reaction diffusion problems in biological systems \cite{Nelson9708071}
as well as advective diffusion in random media \cite{Zoo}.
Interestingly, numerical evidence \cite{HatanoPRL77} suggests the
existence of a localization-delocalization transition of this 2-d
system as a function of the driving potential, to be contrasted with
Hermitian theories for which all wavefunctions are expected to be
localized in $d \leq 2$ \cite{LeeRMP57}.  The nature of the critical
states is however still an open issue.

It is our purpose here to characterize the distribution of critical
wavefunctions for the whole class of non-Abelian random vector
potential models (defined on the group $SU(N)$), by achieving an {\it
exact} calculation of their {\it full} multifractal scaling function
$\tau(q)$.  We do this by reducing the problem to the thermodynamics
of a particle in a random potential used in the Abelian sector
\cite{ChamonPRL77,CastilloPRB56}.  At the fixed point of the theory
located at the limit of infinitely strong disorder, we find that the
multifractal spectrum $\tau(q)$ for $N$ flavours of Dirac
fermions in a non-Abelian random vector potential is
\begin{eqnarray}
\tau(q) = \left\{ \begin{array}{cc}
(q-1)(2-\frac{N-1}{N^2} q) & |q| \leq \frac{\sqrt{2}N}{\sqrt{N-1}} \\
2q (1-\frac{\sqrt{N-1}}{\sqrt{2}N} {\rm sgn}(q)) & |q| >
\frac{\sqrt{2}N}{\sqrt{N-1}} \end{array} \right.
\label{tauN}
\end{eqnarray}
for $N = 2, 3,...$, showing that zero-energy wavefunctions are always
delocalized.  This is (as was first alluded to in \cite{MudryNPB466})
in contrast to the Abelian case \cite{ChamonPRL77,CastilloPRB56}, for
which a sufficiently strong disorder localizes the wavefunctions.

In order to set the stage for the calculation in the non-Abelian case,
let us briefly recall the procedure used in \cite{ChamonPRL77,CastilloPRB56}
to solve the 
Abelian problem.  The Hamiltonian for Dirac fermions interacting with
an Abelian RVP is $\hat{{\cal H}}_A = \not \! \partial + i \not \! \!
A$, where the Pauli matrices are used as the
2-d Euclidean Dirac matrices, i.e. $\not \! \! A = A_{\mu}
\sigma_{\mu}, ~\mu =1,2$.  The first step 
is to find an explicit
expression for the zero-energy wavefunction for each 
realization of the disorder.  In the gauge where $A_{\mu} =
\epsilon_{\mu \nu} \partial_{\nu}\phi$, it can be checked
that $\psi({\bf x}) = e^{-\phi({\bf x})}$ obeys the zero energy Dirac
equation \cite{AharonovPRA19}.  The modulus of the normalized
wavefunction is then 
given by $|\Psi({\bf x})|^2 = e^{-2\phi({\bf x})}/Z(1)$.  
The disorder ensemble averaging on the RVP is
made over a Gaussian white-noise distribution for the
vector potential, i.e. 
\begin{eqnarray}
P[\phi] \propto e^{-S[\phi]}, \hspace{0.7cm} 
S[\phi] = \frac{1}{2 \bar{g}_A} \int d^2 x (\partial_{\mu}
\phi ({\bf x}))^2,
\label{abeliandistribution}
\end{eqnarray}
where $\bar{g}_A$ is the disorder strength.  The key consists in
interpreting $Z(q)$ as a partition 
function for a single particle in
a random site potential $V({\bf x}) = 2 \phi({\bf x})$, in other
words to define the partition function and free
energy as $Z(\beta) = \sum_{\bf x} e^{-\beta V({\bf x})}$ and $F(\beta)
= - \frac{1}{\beta} \ln Z(\beta)$, where the sum is taken over the
$(L/a)^2$ sites of the regularized system. 
By defining a microcanonical partition function (density of states)
$\Omega(E)$ counting the number of energy states in a window of width
$W$ around $E$, 
\begin{eqnarray}
\Omega(E) = \sum_{\bf x} \delta_W (E-V({\bf x})),
\label{microcanonical}
\end{eqnarray}
where $\delta_W (E) = e^{-E^2/2W^2}$, the partition
function can be expressed in terms of the Laplace transform $Z(\beta)
= \int \frac{dE}{W} \Omega(E) e^{-\beta E}$.  Since the disorder
distribution 
(\ref{abeliandistribution}) is a simple Gaussian, it is possible to
calculate directly the average $\langle \Omega(E)
\rangle = \int {\cal D} \phi P[\phi] \Omega(E)$ which depends solely
on the two-point function 
$\langle \phi({\bf x}) \phi ({\bf y}) \rangle = -\frac{1}{2q_{\rm c}^2} \ln
\left\{ \frac{|{\bf x} -{\bf y}|^2 + a^2}{L^2} \right\}$
where $q_{\rm c} = \sqrt{2\pi/{\bar{g}_A}}$. The number of
thermodynamic degrees of freedom being $D = 2 \ln(L/a)$, 
we can define the intensive width, energy and free energy as $w = W/D,
e=E/D, f_0 (\beta) = 
F(\beta)/D$.  The direct calculation \cite{CastilloPRB56} yields
\begin{eqnarray}
\langle \Omega(e) \rangle \sim w \sqrt{\frac{D}{4\pi}} q_{\rm c}
e^{D(1-e^2q_{\rm c}^2/4)} .
\label{disorderaveragedDOS}
\end{eqnarray}
We see that $\langle \Omega(e) \rangle$ has the property of
vanishing for $|e| > e_{\rm c} = 2/q_{\rm c}$ in the thermodynamic limit $D
\rightarrow \infty$, and of
diverging exponentially for $|e|<e_{\rm c}$.  Moreover, it was explicitly
shown in \cite{CastilloPRB56} that in the same limit, for a given
realization of the 
disorder, the energy levels fall within the interval $|e|<e_{\rm c}$ with
probability $1$.  
The entropy density was also argued to be selfaveraging, i.e. quenched and
annealed averages coincide in this region.   

Since there are no states with $e < -e_{\rm c}$, the system undergoes a
freezing transition.  The free energy can be computed from basic
thermodynamic relationships \cite{CastilloPRB56}, and the function
$\tau(q)$ is obtained using the 
correspondence  $\tau(q) = 2q
\lim_{a/L \rightarrow 0} [f_0(q) - f_0 (1)]$.  
In the weak disorder regime defined by $q_{\rm c} > 1$, it is given by
\cite{ChamonPRL77,CastilloPRB56}  
\begin{eqnarray}
\tau(q) = \left\{ \begin{array}{lc}
2(1-\frac{{\rm sgn}(q)}{q_{\rm c}})^2 q, & |q| > q_{\rm c} \\
2(1-\frac{1}{q_{\rm c} ^2}q)(q-1), & |q| \leq q_{\rm c}, \end{array}
\right. 
\label{tau(q)}
\end{eqnarray}
whereas in the strong disorder regime defined by $q_{\rm c} \leq 1$, 
\begin{eqnarray}
\tau(q) = \left\{ \begin{array}{lc}
\frac{4}{q_{\rm c}}(q-|q|), & |q| > q_{\rm c} \\
-2(1-\frac{q}{q_{\rm c}})^2, & |q| \leq q_{\rm c}, \end{array} \right. 
\end{eqnarray}
i.e. the parabolic law completely breaks down.  This spectrum signals
the localization of the wavefunction for strong enough disorder
($\tau(q)$ vanishes for positive $q$).

Let us now turn to the non-Abelian construction.  The principles of
our procedure will
parallel the Abelian one, in needing to find explicit
expressions for the zero-energy wavefunctions and an
effective Gaussian-like field theory for the wavefunction modulus.
This last step will be achieved by taking full advantage of the
underlying group structure of the theory in order to extract only the
elements essential to multifractality.

We thus consider a Fermi surface whose low-energy excitations are
effectively described by $N$ different flavours
of Dirac fermions. The ($2N \times 2N$) Hamiltonian is   
\begin{eqnarray} 
\hat{\cal H}_N = {\bf
I}~ \otimes \not \!  \partial + i \not \! \! A ,
\end{eqnarray} 
where ${\bf I}$ is
the $N \times N$ unit matrix acting in the fermion flavour space.  The $N
\times N$
matrices $A_{\mu}$ belong to the fundamental representation of the
$su(N)$ algebra $(A_{\mu} = A_{\mu}^a t^a, a=1,..,N^2-1)$, where $t^a$
are the generators of $su(N)$ in the fundamental representation.  We
again use Pauli matrices as our 2-d Euclidean
Dirac matrices, i.e. $\not \! \! A = A_{\mu} \otimes \sigma_{\mu}, \mu
=1,2$.  It is convenient to separate the theory
into chiral sectors, for which we define
\begin{eqnarray}
 A_{\pm} ({\bf x}) = A_1({\bf x}) \pm i A_2({\bf x}) = i \partial_{\pm}
g_{\pm} ({\bf x}) g_{\pm}^{-1}({\bf x}), 
\end{eqnarray}
with the reality condition $g_-^{-1} ({\bf x}) = g_+^{\dagger}
({\bf x})$.
The fields $g_{\pm} ({\bf x})$ are 
elements of the (noncompact) complexified group $SU^C(N)$ (obtained
from $SU(N)$ by letting the generating coefficients take values in
$\Bbb{C}$ instead of $\Bbb{R}$).

Zero energy eigenfunctions can be easily
written down by decoupling the wavefunction from the vector fields
with a non-Abelian gauge-like transformation:
\begin{eqnarray}
\Psi_0({\bf x}) = \left( \begin{array}{c} ~~g ({\bf x}) D^+ \\
{{g^{\dagger}}^{-1}}({\bf x}) D^-
\end{array} \right),
\end{eqnarray}
where each of the two entries is composed of the matrix $g ({\bf x})$
multiplying $N$-dimensional column vectors $D^{\pm}$ whose components
are entire functions of $z,\bar{z}$ respectively.  These obey the zero
eigenvalue Dirac equation
\begin{eqnarray}
\hat{\cal H}_N \Psi_0({\bf x}) = 0.
\end{eqnarray}
Note that this is just a convenient rewriting of the wavefunctions
introduced in \cite{Mudry9712103}.  Choosing constant $\{D^{\pm}\}$,
the modulus of these eigenfunctions can be written as
\begin{eqnarray}
|\Psi_0({\bf x})|^2 = {D^+}^{\dagger}_a h_{ab}({\bf x}) D^+_b +
{D^-}^{\dagger}_a {h}^{-1}_{ab}({\bf x}) D^-_b,
\label{norm}
\end{eqnarray}
where we have defined the new matrices $h ({\bf x}) = g^{\dagger}({\bf
x}) g({\bf x})$.  These belong
to the coset $SU^C(N)/SU(N)$, as can be seen from their
invariance under the left-multiplication of $g({\bf x})$ by $u({\bf x}) \in
SU(N)$.  

The disorder averaging is here again made over a Gaussian white-noise
distribution for the vector potential:
\begin{eqnarray}
P[A] \propto \exp{\left\{-\frac{1}{\bar{g}_B} \int d^2 x ~tr ~A_{\mu}^2({\bf
x})\right\}}.
\end{eqnarray}
A striking fact about the NARVP model is that it is a theory which
does {\it not} require the use of either the replica or SUSY
approaches to perform the disorder averaging for zero-energy
eigenfunctions \cite{Bernardh9509137}.  Moreover, the NARVP theory is
known to have a conformally-invariant fixed
point at infinite disorder strength  \cite{MudryNPB466}.  In this
limit, the correct 
formulation of the disorder ensemble averaging leads to a functional
integral over the coset space $SU^C(N)/SU(N)$ with a
Wess-Zumino-Novikov-Witten (WZNW) action on level $k=-2N$ for the
field $h({\bf x})$ \cite{MudryNPB466,CauxPRL80}.  The crucial
point is that 
$h({\bf x})$, which is by construction a primary field of the coset
WZNW model, is precisely the field that appears in (\ref{norm}).

We will use the free field representation of this WZNW model, which
can effectively be found from \cite{GerasimovIJMPA5}.  Omitting the
specifics, we here only state that the result is a
theory for $r$ free bosonic fields which we arrange in a vector
$\boldsymbol{\phi}$ ($r$ is the rank of the algebra, which is $N-1$ in
our case) together with residual fields.  We find
\begin{eqnarray}
S_N = \frac{1}{8\pi} \int d^2 x \left[ N (\partial_{\mu}
\boldsymbol{\phi},\partial_{\mu} \boldsymbol{\phi}) + 2
(\boldsymbol{\rho},\boldsymbol{\phi}){\cal R} \right]
+ S_{\rm res},
\label{freefieldrep}
\end{eqnarray}
where ${\cal R}$ is the curvature (which can be interpreted as a
background charge in the Dotsenko-Fateev construction, and is specific
to the non-Abelian problem), $(.,.)$
represents the scalar 
product in the root space, and $\boldsymbol{\rho}$ is the Weyl vector
(i.e. the half-sum of the positive roots of the algebra).
$S_{\rm res}$ is an action for the residual fields in the 
representation, which are uncoupled from the set $\boldsymbol{\phi}$.  Since
they will not influence the multifractality of the wavefunction,
we can safely discard $S_{\rm res}$.  This free field representation has
highest-weight primary fields \cite{GerasimovIJMPA5}
\begin{eqnarray}
W({\bf x}) = e^{(\boldsymbol{\lambda},\boldsymbol{\phi}({\bf x}))},
\hspace{0.5cm} (\boldsymbol{\alpha}_i, \boldsymbol{\lambda}) =
\delta_{i1}, \hspace{0.5cm} i=1,...,r 
\label{highestweightfield}
\end{eqnarray}
where $\{ \boldsymbol{\alpha}_i \}$ is the set of simple
roots, and $\boldsymbol{\lambda}$ is the first fundamental weight,
defined by the above properties.  

We know (see (\ref{norm})) that the modulus of the wavefunction
involves the matrix field
$h({\bf x})$, which is the primary field of the WZNW model.
We choose the sets $\{D^{\pm}\}$ in such a way
that the highest-weight field (\ref{highestweightfield})
appears in the modulus (this can always be done, since the WZNW model
is invariant under rotation of the field by a constant group element).
The problem is then greatly simplified by decomposing the vector
$\boldsymbol{\phi}$ in 
the (non-orthonormal) basis of the simple roots.  The primary field
(wavefunction modulus) takes in that case the very convenient form of an
exponential of {\it only one} component of the vector field
$\boldsymbol{\phi}$, i.e. 
\begin{eqnarray}
\boldsymbol{\phi} = \sum_{i=1}^r \phi_i \boldsymbol{\alpha}_i,
\hspace{1.0cm} W({\bf x}) = e^{\phi_1 ({\bf x})} .
\end{eqnarray}
The problem is that the different components of the vector field
$\boldsymbol{\phi}$ are still dynamically coupled, since simple
roots are not orthogonal. 
Their scalar product is given by the Cartan matrix
$(\boldsymbol{\alpha}_i, \boldsymbol{\alpha}_j) = A_{ij}, ~~A_{ij} = 2
\delta_{ij} - \delta_{i,j+1} - \delta_{i,j-1}$, 
and the kinetic term in (\ref{freefieldrep}) is 
\begin{eqnarray}
(\partial_{\mu} \boldsymbol{\phi},\partial_{\mu} \boldsymbol{\phi}) =
\sum_{i,j =1}^r A_{ij} 
(\partial_{\mu} 
\phi_i)(\partial_{\mu} \phi_j).
\end{eqnarray}
Since only the field $\phi_1$ appears in the wavefunction modulus, we
wish to decouple it from the set $\{ \phi_i, i=2,...,r \}$.  This can
be done by a simple nonsingular linear transformation (a Harish-Chandra
decomposition),
\begin{eqnarray}
\boldsymbol{\phi} ({\bf x}) = {\bf P} \boldsymbol{\phi}^{\prime} ({\bf x}),
\label{P}
\end{eqnarray}
with $P_{1j} = \frac{N-j}{N-1}$, $P_{jj} = 1$ and $P_{ij} = 0$ for $i
\neq 1, j \neq i$.  Thus, besides the unit diagonal, only the 
first column of ${\bf P}$ is taken to have nonvanishing elements.
Moreover, $\phi_1$ is left unchanged by the transformation (\ref{P}).   
The matrix ${\bf P}^T {\bf A} {\bf P}$ then becomes
\begin{eqnarray}
{\bf P}^T {\bf A} {\bf P} = \left( \begin{array}{ccccc}
C_N & 0 & 0 & 0 & .. \\
0 & 2 & -1 & 0 & .. \\
0 & -1 & 2 & -1 & .. \\
.. & .. & .. & .. & .. \end{array} \right) 
\end{eqnarray}
with $C_N = 2 - P_{12} = \frac{N}{N-1}$.  Using this transformation in
(\ref{freefieldrep}), we achieve our main objective:  the wavefunction
amplitude has completely decoupled from the set $\{ \phi_i
\}, i=2,...,r$, which means that all information about multifractality
is contained in the free field action (we have defined for convenience
the new field $\tilde{\phi} = -\frac{1}{2}\phi_1$)
\begin{eqnarray}
S_N [\tilde{\phi}] = \int d^2 x \left[ \frac{N^2}{2\pi(N-1)}
(\partial_{\mu} \tilde{\phi})^2 - \frac{N}{4\pi} ~{\cal R}
\tilde{\phi} \right]
\label{nonabdistribution}
\end{eqnarray}
representing the weight of the disorder distribution through
$P[\tilde{\phi}] \propto e^{-S[\tilde{\phi}]}$ as far as the
wavefunction modulus $|\Psi_0({\bf x})|^2 = e^{-2\tilde{\phi}({\bf
x})}/Z(1)$ is concerned.  This is one of the central points of our
paper.  

The action (\ref{nonabdistribution}) is a Gaussian Field Theory
with curvature term, and can be viewed as the non-Abelian version of
(\ref{abeliandistribution}).  We can thus perform a calculation for
$\langle \Omega(E) \rangle$ similar to the one in \cite{CastilloPRB56}
(see eq. (\ref{disorderaveragedDOS})), for which we find 
\begin{eqnarray}
\langle \Omega(E) \rangle = W \sum_{\bf x}
\frac{\exp\left\{-\frac{E^2[1-2 c({\bf x})]^2}{2[W^2 + 4 G_0]}
\right\}}{[W^2 + 4 G_0]^{1/2}}, 
\label{OmegaN}
\end{eqnarray}
where $c({\bf x}) = \frac{N}{4\pi E} \int d^2y
\sqrt{g({\bf y})} {\cal R}({\bf y}) G_0({\bf y},{\bf x})$ and $G_0 =
G_0({\bf x},{\bf x})$ is the cutoff limit of the regularized Green's
function for the Laplacian.  We have written $c({\bf x})$ as an
integral defined over a general manifold with metric $g_{\mu \nu}({\bf
x})$, so the square root of the metric's determinant appears in the
measure.  For convenience, we take the topology of our system to be
that of the sphere, and concentrate all the curvature on a single
point that we send to infinity.  In other words, we take $\sqrt{g({\bf
x})} {\cal R}({\bf x}) = 8 \pi \lim_{{\bf B} \rightarrow \infty}
\delta ({\bf B} - {\bf x})$.  The   
short-distance limit of the regularized Green's function then becomes  
$G_0({\bf x},{\bf y}) = - 
\frac{N-1}{4N^2} \ln \left\{ \frac{|{\bf x} -{\bf y}|^2 + a^2}{L^2}
\right\}$.
We find that the terms in (\ref{OmegaN}) involving $c({\bf x})$ are
subdominant in 
the thermodynamic limit $D \rightarrow \infty$, and that the proper
expression for the disorder-averaged number of
states in an energy interval $\omega$ around $e$ (neglecting the width
$W$ as compared to terms of order $\ln (L/a)$) becomes
\begin{eqnarray}
\langle \Omega (e) \rangle \approx \omega \frac{N}{\sqrt{N-1}} \sqrt{D}
\exp\{D(1-e^2/e_{\rm c}^2)\},
\end{eqnarray}
with the critical energy now taking the value $e_{\rm c} =
\sqrt{2(N-1)}/N$.  This expression has exactly the same form as in the
Abelian case \cite{CastilloPRB56}, and we can thus use precisely the
same thermodynamic-like reasoning to obtain $\tau(q)$.
We immediately see that $q_{\rm c} = 2/e_{\rm c} =
\frac{\sqrt{2}N}{\sqrt{N-1}}$ for the non-Abelian case, and that the
full spectrum, in analogy with (\ref{tau(q)}), is given by
(\ref{tauN}).  $q_{\rm c}$ is greater than 
one for every $N$, and we are thus always in the ``weak disorder''
regime of the Abelian solution, for which the logarithm of the
wavefunction normalization factor $Z(1)$ is selfaveraging and the
wavefunctions are delocalized.  Delocalization of 2-d Dirac fermions
can also take place with other forms of disorder (see for a recent
example \cite{ZieglerPRL80}).    

The termination that we have obtained here
is even more stringent than conjectured in \cite{CauxPRL80}:
there, arguments were given for a termination of $\tau^*(q)$
(calculated with {\it un}normalized wavefunctions) in 
general before $q \sim N$, while the case $N=2$ was explicitly
shown to have terminated between $q=2$ and $q=3$.  Here, for $N=2$, we
find $q_{\rm c} = 2\sqrt{2}$, which is consistent with the results coming
from the Logarithmic Conformal Field Theory \cite{CauxPRL80}.

In conclusion, we have studied the distribution characteristics of the
critical wavefunctions of multiflavour Dirac fermions in a non-Abelian
random vector potential, by calculating
exactly their full multifractal spectrum for any number of flavours.
This was achieved by building the relevant Gaussian field theory for
wavefunction amplitudes and using the mapping to the thermodynamics
of a particle in a random potential as in the Abelian case.  Our
multifractal spectrum confirms that zero-energy wavefunctions are
delocalized in the limit of infinite disorder strength, and yields
further evidence for the existence of such types of delocalized states
in 2-d non-Hermitian Quantum Mechanics.

The author is greatly indebted to A. M. Tsvelik, N. Taniguchi,
F. Essler, J. T. Chalker, J. L. Cardy, I. I. Kogan, O. Soloviev,
J.L. Jacobsen and especially C. Mudry for many suggestions and
comments, and acknowledges support from NSERC (Canada).

\widetext

\end{document}